\documentclass[aps,prl,showpacs,twocolumn,floats,floatfix]{revtex4}
\usepackage{amsfonts,amssymb,amsmath}
\usepackage{graphics}
\usepackage{longtable}
\usepackage{subfigure}
\usepackage{epsfig}

\def\la{\left\langle}
\def\ra{\right\rangle}
\begin{document}
\title{Dynamics of Passive-Scalar Turbulence} 
\author{Dhrubaditya Mitra} 
\author{Rahul Pandit}
\altaffiliation[Also at ]{Jawaharlal Nehru Centre For Advanced
Scientific Research, Jakkur, Bangalore, India}
\affiliation{Centre for Condensed Matter Theory, 
Department of Physics, Indian Institute of Science, 
Bangalore 560012, India}
\begin{abstract}
We present the first study of the dynamic scaling or multiscaling
of passive-scalar turbulence. For the Kraichnan version of 
passive-scalar turbulence we show analytically, in both Eulerian and 
quasi-Lagrangian frameworks, that simple dynamic scaling is obtained 
but with different dynamic exponents.  By developing the multifractal
 model we show that dynamic multiscaling occurs in passive-scalar 
turbulence only if the advecting velocity field is itself 
multifractal.  We substantiate our results by detailed numerical 
simulations in shell models of passive-scalar advection.
\end{abstract}
\maketitle
Important advances have been made over the past decade in 
understanding the statistical properties of the turbulence of 
passive scalars, e.g., pollutants, and passive vectors, e.g.,
weak magnetic fields, advected by a fluid~\cite{fal01}.
If the advecting velocity is stochastic and of the Kraichnan 
type~\cite{kra68}, then, in some limits, it can be established 
analytically that passive-scalar and passive-vector turbulence 
show {\it anomalous scaling} or {\it multiscaling} of structure 
functions.  These are the only turbulence problems for which such 
multiscaling can be proven analytically, so it is important to 
use them as testing grounds for new ideas about multiscaling in 
turbulence. We examine the dynamic-scaling properties of 
passive-scalar turbulence in the light of the 
recent systematization of dynamic multiscaling in fluid 
turbulence~\cite{mit04}. 

Our principal results, obtained analytically and numerically, 
illustrate important principles that appear, at first sight, to 
be surprising. We find, e.g., that the dynamic exponents depend 
via bridge relations only on the equal-time scaling exponents of 
the {\it velocity} field. Thus, even though equal-time structure 
functions for the passive-scalar problem display multiscaling, we 
show analytically that they exhibit {\it simple dynamic scaling} 
if the advecting velocity is of the Kraichnan type. Our study 
of the Kraichnan model yields the only analytical results
obtained so far for time-dependent structure functions in any type 
of turbulence.  Dynamic multiscaling is obtained only if the 
advecting velocity field is itself intermittent as we demonstrate
numerically. 

The quest for a statistical characterization of turbulence begins 
most often with the equal-time, order-$p$, velocity-$u$ structure 
functions $S^u_p(\ell)$, i.e., the order-$p$ moments of the 
probability distribution functions~(PDFs) of velocity differences 
at the length scale $\ell$.  The equal-time exponents $\zeta^u_p$ 
are defined by $S^u_p(\ell) \sim \ell^{\zeta^u_p}$, valid for the 
inertial range $\eta_d \ll \ell \ll L $, where $\eta_d$ is the 
dissipation scale and $L$ the length at which energy is pumped in. 
Kolmogorov's simple scaling~\cite{kol41}(K41) yields 
$\zeta_p^{u,K41} = p/3$, but subsequent work~\cite{Fri96} 
suggests significant corrections for $p > 3$ and 
{\it multiscaling} with $\zeta^u_p$ a nonlinear, convex, 
monotonically increasing function of $p$.  The generalization of 
such multiscaling to {\it dynamic multiscaling} is subtle and has
been elucidated only recently~\cite{lvo97,hay98,mit03,mit04}: It 
is expected that Eulerian-velocity time-dependent structure 
functions~\cite{kan99} lead to trivial dynamic scaling with all 
Eulerian~(${\cal E}$) dynamic exponents $z_p^{u,\cal{E}} = 1$.  
Nontrivial dynamic exponents $z^u_p$ can be obtained~\cite{mit04} 
from dynamic-multiscaling ans\"atze of the form 
$\tau^u_p \sim \ell^{z^u_p}$, where the times $\tau^u_p$ are 
extracted from {\it Lagrangian} or {\it quasi-Lagrangian} 
time-dependent structure functions.  A generalization of the 
multifractal formalism~\cite{Fri96} to the case of time-dependent 
structure functions~\cite{lvo97,mit03,mit04} shows that dynamic 
and equal-time multiscaling exponents must be related by 
{\it bridge relations}; and these bridge relations depend 
crucially~\cite{mit04} on how time scales are extracted:  In 
particular, time scales extracted via time derivatives of 
time-dependent structure functions are different from those 
obtained from integrals.  

 The advection--diffusion equation for the passive scalar 
field $\theta({\bf x},t)$ at point ${\bf x}$ and time $t$ is 
\begin{equation}
\partial_t\theta +  u_i \partial_i \theta 
  = \kappa \partial_{ii} \theta + f_{\theta},   
\label{eq:pas:passcal}
\end{equation}
where $\kappa$ is the passive-scalar diffusivity and $f_{\theta}$
 an external force. The advecting velocity ${\bf u}$ should be 
obtained from solutions of the Navier--Stokes equation, but, to 
investigate equal-time multiscaling of passive-scalar structure 
functions, it has proved fruitful to use the {\it Kraichnan 
ensemble} in which each component of ${\bf u}$ is a zero-mean, 
delta-correlated Gaussian random variable with  
\begin{equation}
\la u_i({\bf x},t)u_j({\bf x}+{\bf r},t^{\prime}) \ra 
            = 2D_{ij}({\bf r})\delta(t-t^{\prime}).
\label{eq:pas:ucor}
\end{equation} 
The Fourier transform of $D_{ij}$ has the form
\begin{equation}
{\tilde D}_{ij}({\bf q}) \propto 
          \left(q^2 +\frac{1}{L^2}\right)^{-\frac{d+\xi}{2}} 
          e^{-\eta q^2}
		  \left[\delta_{ij} -\frac{q_iq_j}{q^2} \right],
\label{eq:pas:D}
\end{equation}
where ${\bf q}$ is the wave-vector, $d$ the spatial dimension, 
$\eta$  the dissipation scale, $L$ the large forcing scale, 
and $\xi$ a parameter.  The term inside square brackets  assures 
incompressibility.  In real space, if we write 
$D_{ij}({\bf r}) \equiv D^0\delta_{ij} - 
                    \frac{1}{2}d_{ij}({\bf r})$
and take the limits $L \to \infty$ and $\eta \to0$, we get 
\begin{equation}
d_{ij} = D_1 r^{\xi}\left[ (d-1+\xi)\delta_{ij} - 
                 \xi\frac{r_i r_j}{r^2} \right],
\label{eq:pas:dij}
\end{equation}
with $D_1$ a normalization constant.  For $0 < \xi < 2$ this model
shows multiscaling of equal-time, order-$p$, passive-scalar 
structure functions~\cite{fal01}.  The constant 
$D^0 \equiv  2\int_0^{\infty}{\tilde D}_{ij}({\bf q})d^dq 
      \propto O(L^{\xi})$
diverges in the limit $L \to \infty$.  The external force 
$f_{\theta}$ is also a zero-mean, Gaussian random variable
which is white-in-time with the variance
$\la f_{\theta}({\bf x},t)f_{\theta}({\bf y},t^{\prime}) \ra = 
     {\mathcal C}\left(\frac{\mid {\bf x} - {\bf y}\mid}{L} \right)
        \delta(t-t^{\prime})$, 
where the function ${\mathcal C}(x/L)$ is confined to large length
scales.  Moreover, $f_{\theta}$ and ${\bf u}$ are statistically 
independent.

The quasi-Lagrangian transformation of any Eulerian field 
$\psi({\bf x},t)$ is defined by 
${\hat \psi}({\bf x},t) \equiv 
      \psi[{\bf x}+{\bf R}(t;{\bf r}_0,0),t]$,
where ${\bf R}(t;{\bf r}_0,0)$	is the Lagrangian trajectory 
passing through ${\bf r}_0$ at $t=0$~\cite{bel87}. Thus the 
quasi-Lagrangian version of Eq.~(\ref{eq:pas:passcal}) is
\begin{equation}
\partial_t{\hat \theta} + 
\left[{\hat u}_i - {\hat  u}_i(0)\right] \partial_i {\hat \theta}
  = \kappa \partial_{ii} {\hat \theta} + {\hat f}_{\theta}.   
\label{eq:pas:qlpasscal}
\end{equation}
The equal-time multiscaling properties of passive-scalar structure 
functions in the Kraichnan model, well understood in the Eulerian 
framework, remain unchanged in the quasi-Lagrangian 
framework~\cite{fal01}.  We concentrate on the time-dependent 
structure functions 
\begin{equation}
{\mathcal F}^{\phi}_p({\bf r},\{t_1,\ldots,t_p\}) \equiv 
        \la [\delta \phi({\bf x},t_1,{\bf r}) \ldots  
                 \delta \phi({\bf x},t_p,{\bf r})] \ra ,  
\label{eq:pas:deltas} 
\end{equation}
where $\phi$ is $\theta$ or ${\hat \theta}$ in the Eulerian or 
quasi-Lagrangian frameworks, respectively,  the angular brackets 
denote an average over the PDFs of ${\bf u}$ and $f_{\theta}$, and 
$\delta \phi({\bf x},t,{\bf r}) \equiv \phi({\bf x}+{\bf r},t)
 -\phi({\bf x},t).$ 
Dimensional analysis yields the $r$-dependent characteristic time  
${\mathcal T}(r) \sim \frac{r^2}{D_{ij}(r)} \sim r^{2-\xi}$,
whence $z^{\hat \theta}_p = 2 - \xi$ for all $p$. Like the K41 
result $z^u_p=2/3$ for fluid turbulence, we expect this prediction 
to be valid for the passive-scalar case in Lagrangian or 
quasi-Lagrangian frameworks.  For $p=2$, 
${\mathcal F}^{\phi}_2(r,t) =  2C^{\phi}({\bf 0},t) 
                                  - 2C^{\phi}({\bf r},t)$,
where 
$C^{\phi}({\bf r},t) \equiv 
         \la \phi({\bf x}+{\bf r},t)\phi({\bf x},0) \ra$.
Equations for $C^{\phi}({\bf r},t)$, obtained
from Eqs.~(\ref{eq:pas:passcal})~and~(\ref{eq:pas:qlpasscal}) by  
Gaussian averaging~\cite{Fri96} over the  PDFs of ${\bf u}$ and 
${\bf f}_{\theta}$~\footnote{The equation for $C^{\phi}({\bf r},t)$ 
is closed only for strictly positive $t$, and this is the case we 
consider.}, are (in the $\kappa\to0$ limit of relevance to 
turbulence)
\begin{eqnarray}
\label{eq:pas:Clt}
\partial_t C^{\theta}({\bf r},t) &=& 
     D^0(L)\partial_{ii}C^{\theta}
    \sim L^{\xi}\partial_{ii}C^{\theta}; \\
\partial_t C^{{\hat \theta}}({\bf r},t) &=&
     \left(D^0\delta_{ij} - D_{ij} \right) \partial_{ij}C^{{\hat \theta}} 
    \sim d_{ij}({\bf r})\partial_{ij}C^{{\hat \theta}}.  
\label{eq:pas:qlClt}
\end{eqnarray}
Spatial Fourier transforms of Eqs.~(\ref{eq:pas:Clt}) or 
(\ref{eq:pas:qlClt}) then yield 
$ {\tilde C}^{\phi}({\bf q},t) \sim \exp[-t/\tau^{\phi}({\bf q})]$,
with characteristic time scales $\tau^{\theta} = [D^0(L)q]^{-2}$
[Eq.~(\ref{eq:pas:Clt})] and $\tau^{{\hat \theta}} \sim q^{\xi -2}$
[Eq.~(\ref{eq:pas:qlClt})].  
 The last term of Eq.~(\ref{eq:pas:Clt}) diverges 
as $L\to \infty$,  a signature of the sweeping effect; however, for 
a fixed integral scale $L$, the Fourier transform of  
Eq.~(\ref{eq:pas:Clt}) implies  
$z_2^{\theta}=2$ (cf. $z^{u,\cal E}_p = 1$).  This sweeping 
divergence is removed in Eq.~(\ref{eq:pas:qlClt}), whose
Fourier transform, in the limit $L\to\infty$, gives 
$z^{{\hat \theta}}_2 = 2 - \xi$, which agrees with the dimensional 
prediction. These results are remarkable for two reasons: 
(1) this is the first calculation of dynamic scaling exponents in 
any form of turbulence and (2) this shows explicitly how  
sweeping effects are removed in the quasi--Lagrangian representation.
Similar, but more cumbersome, calculations yield $z^{\theta}_4=2$ and
 $z^{{\hat \theta}}_4 = 2 - \xi$, which we substantiate by our
numerical studies below~\footnote{For such simple scaling the 
integral and derivative time scales of course yield the same 
dynamic-scaling exponent.}.  

A turbulent velocity field obeying the Navier--Stokes equation does 
not have simple statistical properties as in Eq.~(\ref{eq:pas:ucor}),
so we use the multifractal model~\cite{Fri96} 
and its extension to dynamic multiscaling~\cite{mit04}. 
To eliminate the sweeping effect, we consider the
structure functions, namely 
${\mathcal F}^{{\hat \theta}}_p(r,t)$,  namely,
\begin{eqnarray}
 \frac{{\mathcal F}^{{\hat \theta}}_p(r,t)}{({\hat \theta}_L)^p} 
		\propto \int_{{\cal I}}d\mu(h)
 (\frac{r}{L})^{3+ph-D(h)}{\cal G}^{p,h}(\frac{t}{\tau_{p,h}}),
\label{mcalf}
\end{eqnarray}
where ${\hat \theta}$ is assumed to possess a range of universal 
scaling exponents $h\in{\cal I} \equiv (h_{min},h_{max})$. For each 
$h$ in this range, there exists a set 
$\Sigma_h \subset \mathbb{R}^3$ of fractal dimension $D(h)$, 
such that 
$ \frac{\delta{\hat\theta}({{\bf x}},r)}{{\hat \theta}_L} 
\propto (\frac{r}{L})^h $
  for ${\bf x} \in {\Sigma}_h$ ,
with ${\hat \theta}_L$ the passive-scalar variable at the forcing 
scale $L$. ${\cal G}^{p,h}(\frac{t}{\tau_{p,h}})$ has a 
characteristic decay time 
$\tau_{p,h} \sim r/\delta u(r)$~ \footnote{This is 
the simplest ansatz for $\tau_{p,h}$~\cite{mit04}.},
${\cal G}^{p,h}(0) = 1$, and the velocity field is also multifractal 
with a range of universal scaling exponents 
$g\in{\cal I} \equiv (g_{min},g_{max})$.  
Following Ref.~\cite{mit04} we define the order-$p$, degree-$M$, 
{\it integral} time scale
${\cal T}^{{\hat \theta},I}_{p,M}(r) \equiv 
 \biggl[ \frac{1}{{\mathcal S}^{{\hat \theta}}_p(r)}
\int_0^{\infty}{\mathcal F}^{{\hat \theta}}_p(r,t)t^{(M-1)} dt
\biggl]^{(1/M)}$ 
and the {\it derivative} time scale 
$ {\cal T}^{{\hat \theta},D}_{p,M} 
\equiv \biggl[\frac{1}{{\mathcal S}^{{\hat \theta}}_p(r)}
                   \frac{\partial^M}{\partial t^M} 
           {\mathcal F}^{{\hat \theta}}_p(r,t) 
           \biggl|_{t=0} \biggl]^{(-1/M)} ,$
with corresponding dynamic-multiscaling exponents defined via 
${\cal T}^{{\hat \theta},I}_{p,M} \sim 
r^{z^{{\hat \theta},I}_{p,M}}$ and
${\cal T}^{{\hat \theta},D}_{p,M} \sim 
r^{z^{{\hat \theta},D}_{p,M}}$. 
To calculate 
\begin{equation}
{\cal T}^{{\hat \theta},I}_{p,1}(r)
\propto \frac{1}{S^{{\hat \theta}}_p(r)}\int_{{\cal I}}d\mu(h)
 (\frac{r}{L})^{3+ph-D(h)}
\int_0^{\infty}{\cal G}^{p,h}(\frac{t}{\tau_{p,h}}) dt,
\end{equation} 
e.g., we substitute the multifractal form (\ref{mcalf}), 
do the time integral first, and use the scaling ansatz for 
$\tau_{p,h}$ and we obtain 
${\cal T}^{{\hat \theta},I}_{p,1}(r)
\propto r^{1-\zeta^{\theta}_p}
 \langle (\delta\theta)^p(\delta u)^{-1} \rangle $.   
Following Ref.~\cite{jen92}\footnote{As shown in Ref.~\cite{jen92}, 
this assumption is equivalent to assuming strong correlation 
between the flux of the velocity and the passive-scalar varible 
which leads to good agreemnt of $\zeta^{\theta}_p$ with experimental 
data. We have checked that this assumption holds true for our 
shell model simulations for $p,q = 1, \ldots, 6$.}, if we now assume 
that the dominant contribution to 
\begin{equation}
\langle (\delta\theta)^p(\delta u)^{-q} \rangle 
   \approx \langle(\delta\theta)^p \rangle 
           \langle (\delta u)^{-q} \rangle,
\end{equation} 
we have 
$z^{{\hat \theta},I}_{p,M} = 1-|\zeta^u_{-1}|$. 
Similar calculations yield the following more
general bridge-relations
\begin{eqnarray}
z^{{\hat \theta},D}_{p,M} = 1-\frac{\zeta^u_M}{M}, \hspace*{0.4cm} 
z^{{\hat \theta},I}_{p,M} = 1-\frac{|\zeta^u_{-M}|}{M},
\label{eq:pas:zp_pshell}
\end{eqnarray}
which do not depend on $p$. However, this does not mean we have 
simple dynamic scaling. For a velocity field which multiscales, i.e.,
for which $\zeta^u_M/M \ne \zeta^u_1$, we have dynamic multiscaling.
For the Kraichnan model a similar multifractal formalism predicts 
$z^{{\hat \theta},D}_{p,1} = 2 -\xi$. 

We now substantiate our analytical and multifractal results by 
detailed numerical simulations of shell-model analogs of the two 
passive-scalar problems discussed above. For this purpose we 
consider two very similar shell models, A and B, 
of the general form~\cite{wir96,jen92} 
\begin{equation}
[\frac{d}{dt} + \kappa k_m^2] {\hat \theta}_m(t) 
=  i \Phi^{{A/B}}_{m,{\hat \theta} {\hat u}} + \delta_{m,1}f(t).
\label{eq:pas:kshell}
\end{equation}
For  model A, which corresponds to the Kraichnan model, 
$\Phi^A_{m,{\hat \theta} {\hat u}} = 
 [(k_m/2)({\hat \theta}^{\ast}_{m+1}{\hat u}^{\ast}_{m-1} - 
          {\hat \theta}^{\ast}_{m-1}{\hat u}^{\ast}_{m+1}) 
+(-k_{m-1}/2)({\hat \theta}^{\ast}_{m-1}{\hat u}^{\ast}_{m-2}+
         {\hat \theta}^{\ast}_{m-2}{\hat u}_{m-1}) 
+(k_{m+1}/2)({\hat \theta}^{\ast}_{m+2}{\hat u}_{m+1}+
       {\hat \theta}^{\ast}_{m+1}{\hat u}^{\ast}_{m+2})]$.
Here the asterisk denotes complex conjugation, ${\hat \theta}_m$ 
and ${\hat u}_m$ are, respectively, shell-model analogs of the 
Fourier components of the passive scalar and velocity 
in the quasi-Lagrangian framework,   
$k_m = 2^m k_0$ and $k_0 =1/16$.  The shell velocity is a 
zero-mean, Gaussian random complex variable with covariance
$\la {\hat u}_m(t){\hat u}^{\ast}_n(t^{\prime})\ra =
		D_m\delta_{m,n}\delta(t-t^{\prime})$,
$D_m = k_m^{-\xi}$, $f(t)$ is random, Gaussian and white-in-time 
and independent of ${\hat u}_m$. In model B,
$\Phi^B_{m,{\hat \theta}{\hat u}} = 
[k_m({\hat \theta}_{m+1}{\hat u}_{m-1}-
       {\hat \theta}_{m-1}{\hat u}_{m+1}) 
-(k_{m-1}/2)({\hat \theta}_{m-1}{\hat u}_{m-2}+
        {\hat \theta}_{m-2}{\hat u}_{m-1}) 
(-k_{m+1}/2)({\hat \theta}_{m+2}{\hat u}_{m+1}+
   {\hat \theta}_{m+1}{\hat u}_{m+2})]^{\ast}$.
This is the shell-model analog of a passive scalar 
advected by a Navier--Stokes velocity field. 
Here the shell velocity obeys the GOY shell-model 
equation~\cite{Fri96,goy}  
\begin{equation}
[\frac{d}{dt}+\nu k_m^2]{\hat u}_m =  i\Gamma_{m,{\hat u}{\hat u}} 
 						+ \delta_{m,1} f^{{\hat u}},
\label{eq:pas:goy}
\end{equation}
where $\Gamma_{m,{\hat u}{\hat u}} = 
[k_m {\hat u}_{m+1}{\hat u}_{m+2}
        -\delta k_{m-1}{\hat u}_{m-1}{\hat u}_{m+1}  
     -(1-\delta)k_{m-2}{\hat u}_{m-1}{\hat u}_{m-2}]^{\ast}$ 
with $\delta=1/2$. For both models A and B, 
${\hat u}_{-1} = {\hat u}_{0} = {\hat \theta}_{-1} = 
{\hat \theta}_{0} = 0$, and the couplings are limited to 
next-nearest-neighbor shells, hence the dynamic scaling or 
multiscaling properties of time-dependent structure functions 
should be akin to those of quasi-Lagrangian variables~\cite{mit04}
~(so we use ${\hat \theta}_m$ and ${\hat u}_m$).
Furthermore, when $\nu \to 0$ and $f^{\hat u}\to 0$, 
$E^{{\hat \theta}} \equiv 
 \sum_{m=1}^N \mid~{\hat \theta}_m~\mid^2$ is 
conserved, where $N$ is the total number of shells.  We use a weak, 
order-one, Euler scheme associated with the Ito formulation of 
Eq.~(\ref{eq:pas:kshell}) to integrate model A and a second-order
Adams-Bashforth scheme to integrate model B. The different 
parameters used in our simulations, e.g., the 
large-eddy-turnover-time~$\tau_L$,  
are given in Table~(\ref{table:pas:para}).  
The shell-model equal-time passive-scalar structure functions are 
$S_p(m) \equiv \la [{\hat \theta}_m {\hat \theta}_m^{\ast}]^{p/2} \ra
\sim k_m^{-\zeta^{\theta}_p}$.
Model A exhibits equal-time multiscaling for $0 < \xi < 2$; 
we use $\xi = 0.6$ here. The equal-time multiscaling 
exponents for model B are given in the second column of 
Table~\ref{table:pexp}. Our results for the equal-time multiscaling 
exponents~$\zeta^{\theta}_p$ for both models A and B agree well with 
previous studies~(Ref.~\cite{wir96} for model A and 
Ref.~\cite{jen92} for model B).  We now define the order-$p$, 
time-dependent, passive-scalar structure functions for our
shell models: 
\begin{equation}
 F_p(m,t) \equiv \frac{1}{S_p(m)}\la [{\hat \theta}_m(0)
              {\hat \theta}_m^{\ast}(t)]^{p/2} \ra. 
\label{eq:pas:kshFp}
\end{equation}
For model A an analytical calculation yields [cf. 
Eq.~(\ref{eq:pas:Clt})]
\begin{equation}   
F_2(m,t) = S_2(m) \exp[-\frac{1}{4}k_m^{2-\xi}A(\xi)t],
\label{eq:pas:f2kshell}
\end{equation}
where 
$A(\xi) = [2^{(2\xi-2)}+ 2^{-(2\xi-2)}) + (2^{\xi} + 2^{-\xi})
               + (2^{(\xi-2)} + 2^{-(\xi-2)}]$, whence 
$z^{{\hat \theta}}_2=2-\xi$.
Similar relations can be derived for $p\geq4$ but the complexity
increases with $p$.

\begin{figure}[ht]
\begin{minipage}[t]{\linewidth}
\includegraphics[width=0.6\linewidth]{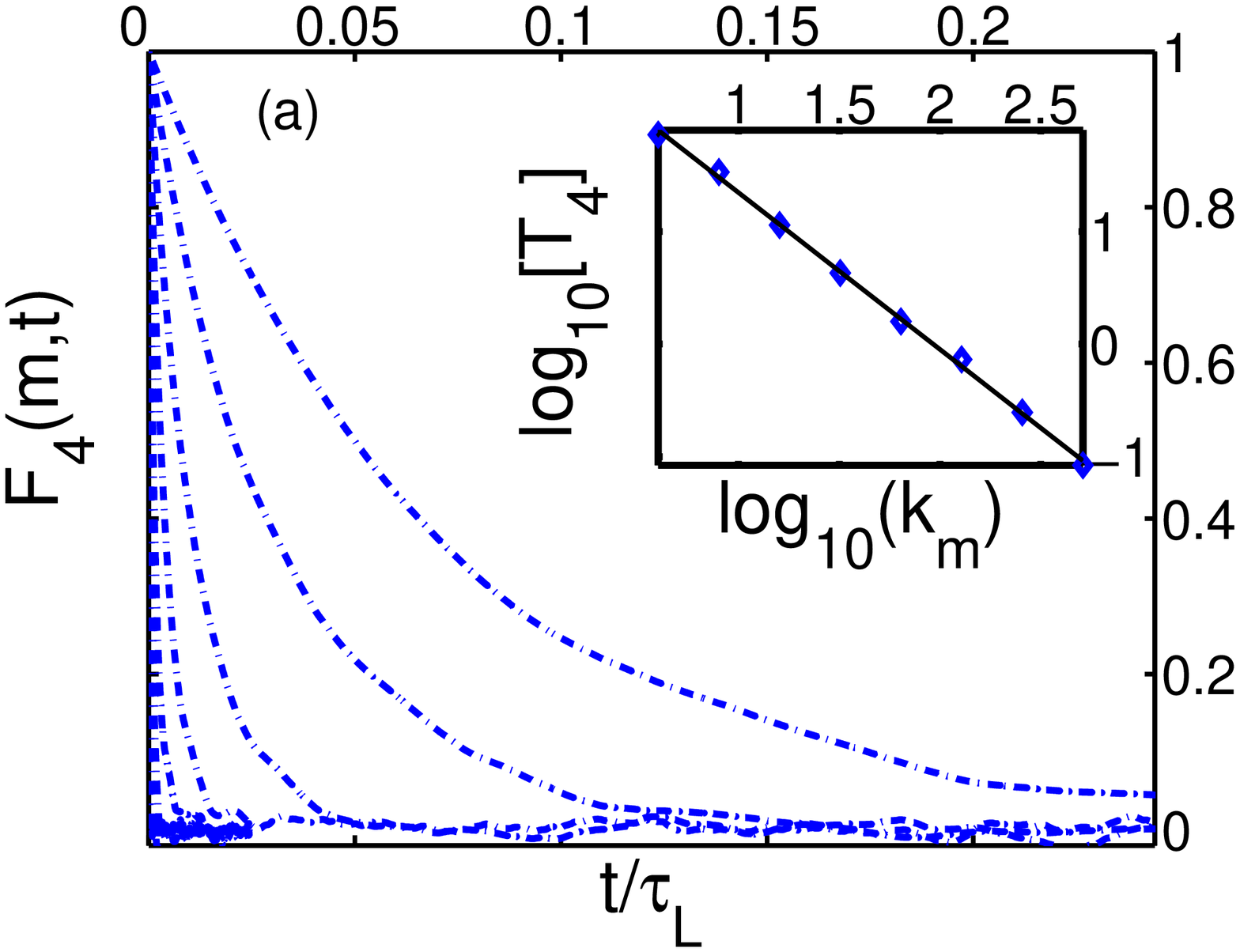}  
\includegraphics[width=0.6\linewidth]{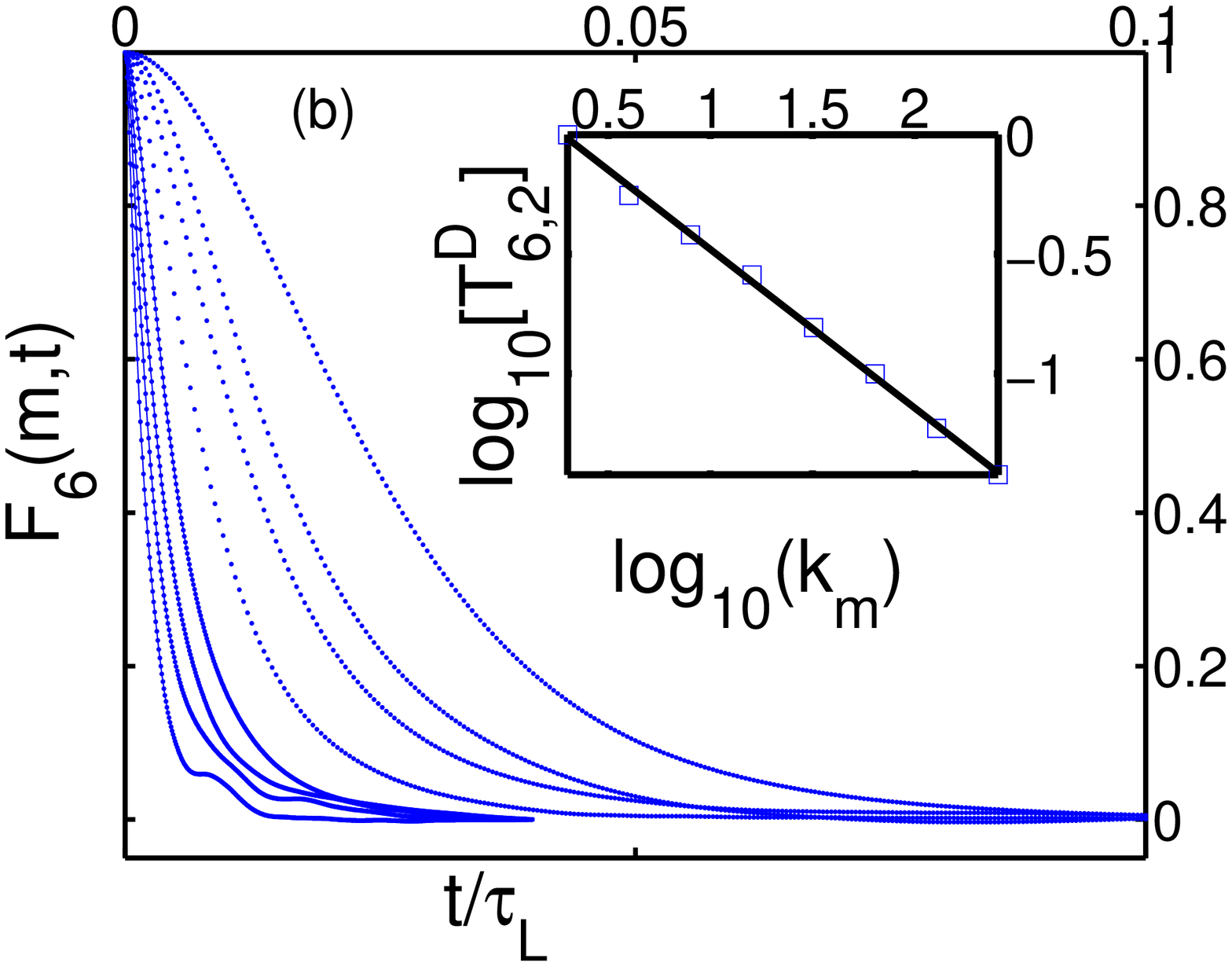}
\end{minipage}
\caption{\small (a) Plot of the order-$4$ time-dependent structure 
function for  model A, with $\xi = 0.6$, versus $t/\tau_L$,
for shells $m=6$ to $13$~(the last few are not clearly visible). 
Exponential functions [like Eq.~(\ref{eq:pas:f2kshell})]
are good approximations for these time-dependent structure functions.
 The inset shows a log-log plot of  $T^{{\hat \theta}}_4(m)$ 
against the wavevector $k_m$; the slope of the straight line 
(least-squares fit) gives 
$z^{{\hat \theta}}_4 \simeq 1.40 \simeq 2-\xi$.  
(b) A plot as in (a) for model B, for $p=6$. The inset shows 
a log-log plot of $T^D_{6,2}$ versus the wavevector 
$k_m$; the straight line is the least-squares fit to the points.}
\label{fig:all}
\end{figure}

For model A we fit an exponential to each of $F_p(m,t)$ up to a 
time $t_{\mu}$, such that 
$\frac{F_p(m,t_{\mu})}{S_p(m)} = \mu$, 
to extract a characteristic decay rate $T_p(m)$.  We use  
$\mu=0.7$; we find that values of $\mu$ from $0.5$ to $0.9$ 
do not change our results significantly. Representative plots for 
$p=4$ are shown in Fig.~(\ref{fig:all}~a). The slopes of these and 
similar plots yield 
$z^{{\hat \theta}}_2=1.400\pm0.005$, 
$z^{{\hat \theta}}_4=1.400\pm005$, 
$z^{{\hat \theta}}_6=1.40\pm0.01$, which clearly support our 
analytical result $z^{{\hat \theta}}_p=2-\xi$ with $\xi = 0.6$.
\begin{table}
\framebox{\begin{tabular}{c|c|c|c|c|c|}
Model&$\kappa$&$\delta t$&$\tau_L$&$T_{tr}$&$T_{av}$\\
\hline
A&$2^{-14}$&$2^{-24}$&$\simeq 2^{24}\delta t$&$5\times 10^4\tau_L$& 
$10^5\tau_L$\\
B&$5\times 10^{-7}$&$10^{-4}$&$10^5\delta t$&$5\times 10^4\tau_L$&
$10^5\tau_L$\\
\end{tabular}}
\caption{The diffusivity $\kappa$, the time-step $\delta t$, and 
the box-size eddy turnover time $\tau_L \equiv 1/k_0u_{rms}$ that 
we use in our simulations of models A and B.  Data 
from the first $T_{tr}$ time steps are discarded so that transients 
can die down. We then average our data for time-dependent structure 
functions for an averaging time $ T_{av}$. For  model A we use 
$\xi = 0.6$. The number of shells $N=22$ for both the models.}
\label{table:pas:para}
\end{table}
\begin{table}
\framebox{\begin{tabular}{c|c|c|c|c|}
order$(p)$ & $\zeta^{{\hat \theta}}_p$&$z_{p,1}^{\theta,I}$ 
&$z_{p,2}^{{\hat \theta},D}$\\
\hline
1& $0.34\pm0.001$ & $0.52\pm0.03$  & $0.63\pm0.03$ \\
2& $0.63\pm0.001$ & $0.53\pm0.03$  & $0.64\pm0.03$ \\ 
3& $0.87\pm0.001$ & $0.56\pm0.005$ & $0.64\pm0.005$\\ 
4& $1.07\pm0.001$ & $0.56\pm0.005$ & $0.64\pm0.01$ \\ 
5& $1.24\pm0.004$ & $0.56\pm0.005$ & $0.64\pm0.01$ \\ 
6& $1.38\pm0.006$ & $0.57\pm0.007$ & $0.64\pm0.02$ \\
\end{tabular}}
\caption{Order$-p\/$ (Column 1) multiscaling exponents for 
$1\leq p \leq6$ from our simulations of model B: equal-time 
exponents $\zeta^{\theta}_p$ (Column 2), integral-scale 
dynamic-multiscaling exponent 
$z^{{\hat \theta},I}_{p,1}\/$ of degree-$1$ 
(Column 3) and  derivative-time dynamic-multiscaling exponents 
$z^{{\hat \theta},D}_{p,2}$ (Column 4) from our calculations. 
We obtain 
(1) $z^{{\hat \theta},I}_{p,1} = 0.56\pm0.005$ from the 
bridge-relation (\ref{eq:pas:zp_pshell}) and 
$\zeta^u_{-1} =-0.44\pm0.005$; 
(2) $z^{{\hat \theta},D}_{p,2} = 0.645\pm0.0001$ from the bridge 
relation and $\zeta^u_2 = 0.709\pm 0.0001$.  
See text for error estimates.}
\label{table:pexp}
\end{table}
For model B the imaginary parts of $F_p(m,t)$ are negligible compared
to their real parts.  Henceforth we consider only the real part. 
We use two different sampling rates, $50\times \delta t$ for 
$4\le m \le 8$ and  $10\times \delta t$ for $9\le m\le 13$, 
respectively.  For extracting  
$T^D_{p,2}$ we extend $F_p(m,t)$ to negative $t$ via 
$F_p(m,-t) = F_p(m,t)$ and use a centered, sixth-order, 
finite-difference scheme to find 
$\frac{\partial^2}{\partial t^2} F_p(m,t) \bigl|_{t=0}$.
A log-log plot of $T^D_{p,2}$  versus $k_m$ now 
yields the exponents $z^{{\hat \theta},D}_{p,2}$ given in 
Table~(\ref{table:pexp}).  A comparison of our results with the 
multifractal prediction of Eq.(\ref{eq:pas:zp_pshell}) is shown in 
Table~(\ref{table:pexp}).  
 We have computed the integral time scale only for $M=1$, i.e.,  
$T^I_{p,1}(m,t_u) \equiv \int_0^{t_u}F_p(m,t) dt  
\sim k_m^{-z^{{\hat \theta},I}_{p,1}}$ \footnote{We have checked that 
$\zeta^u_p$ are well defined here only for $p>-1.84$ so we must have
$M<1.84$ via Eq.~(\ref{eq:pas:zp_pshell}).}. 
In principle we should use $t_u \rightarrow \infty$ but, since it is
not possible to obtain $F_p(m,t)$ accurately for large $t$, we 
select an upper cut-off $t_u$ such that $F_p(m,t_u) = \alpha$, where,
for all $m$ and $p$, we choose $\alpha = 0.7$.  
We have checked that our results do not change if we use 
$0.3 < \alpha < 0.8$.  The slope of a log-log plot of 
$T^I_{p,1}(m)$ versus $k_m$ now yields 
$z^{{\hat \theta},I}_{p,1}$ shown in Table~(\ref{table:pexp}). 
The difference between the dynamic-scaling exponents of integral 
and derivative type~(Columns 3 and 4 of 
Table~(\ref{table:pexp}), respectively) is a clear signature of 
dynamic multiscaling. 

To obtain the  error-bars on the equal-time and dynamic exponents,
we carry out $50$ runs, each averaged over a time $T_{av}$ given
in Table~(\ref{table:pas:para}). We thus obtain $50$ different 
values for each of these exponents. The mean values of these 
$50$~exponents are quoted here~[Table~(\ref{table:pexp})]; and the 
root-mean-square deviation about the mean value yields the 
error estimates shown in Table~(\ref{table:pexp}).
Note that the low-order dynamic structure functions decay
slowly and hence, for a fixed averaging time, low-order dynamic
multiscaling exponents have larger error bars than those
of slightly higher-order exponents.  
 
We have shown that, for  passive-scalar turbulence, simple dynamic 
scaling is obtained if the velocity field is of the Kraichnan type. 
Dynamic multiscaling is obtained if the advecting velocity is 
itself multifractal. The study of dynamic multiscaling in 
passive-vector turbulence is more complicated because of the 
dynamo effect and will be dealt with elsewhere.
The experimental study of such dynamic multiscaling 
remains an important challenge.

We thank A.~Celani, S.~Ramaswamy and M.~Vergassola for discussions
and IFCPAR~(Project no. 2404-2) and CSIR~(INDIA) for financial 
support.

\end{document}